\documentclass{vow2008}

\title{Science with VO tools: the AstroGrid VO Desktop}
\author{Jonathan A. Tedds}
\affil{University of Leicester (AstroGrid/EuroVO/XMM-Newton Survey Science Centre)}

\usepackage{graphicx}
\begin{document}

\keywords{Virtual Observatory, archives, services, data, metadata, registry, catalogs, cross-match techniques, searching, visualization, X-ray, clusters of galaxies}

\maketitle



\begin{abstract}          
We introduce a general range of science drivers for using the Virtual Observatory (VO) and identify some common aspects to these as well as the advantages of VO data access. We then illustrate the use of existing VO tools to tackle multi wavelength
science problems. We demonstrate the ease of multi mission data access
using the VOExplorer resource browser, as provided by AstroGrid (http://www.astrogrid.org) and show how to pass the various results into any VO enabled tool such as TopCat for catalogue correlation. 
VOExplorer
offers a powerful data-centric visualisation for browsing and filtering the
entire VO registry using an iTunes type interface. This allows the user to bookmark their own personalised lists of resources and to run tasks
on the selected resources as desired. We introduce an example of how more advanced querying can be performed to access existing X-ray cluster of galaxies catalogues and then select extended only X-ray sources as candidate clusters of galaxies in the 2XMMi catalogue. Finally we introduce scripted access to VO resources using python with AstroGrid and demonstrate how the user can pass on the results of such a search and correlate with e.g. optical datasets such as Sloan. Hence we illustrate the power of enabling large scale data mining of multi wavelength resources in an easily reproducible way using the VO.
\end{abstract}

\begin{figure*}
\centering
\includegraphics[scale=0.7]{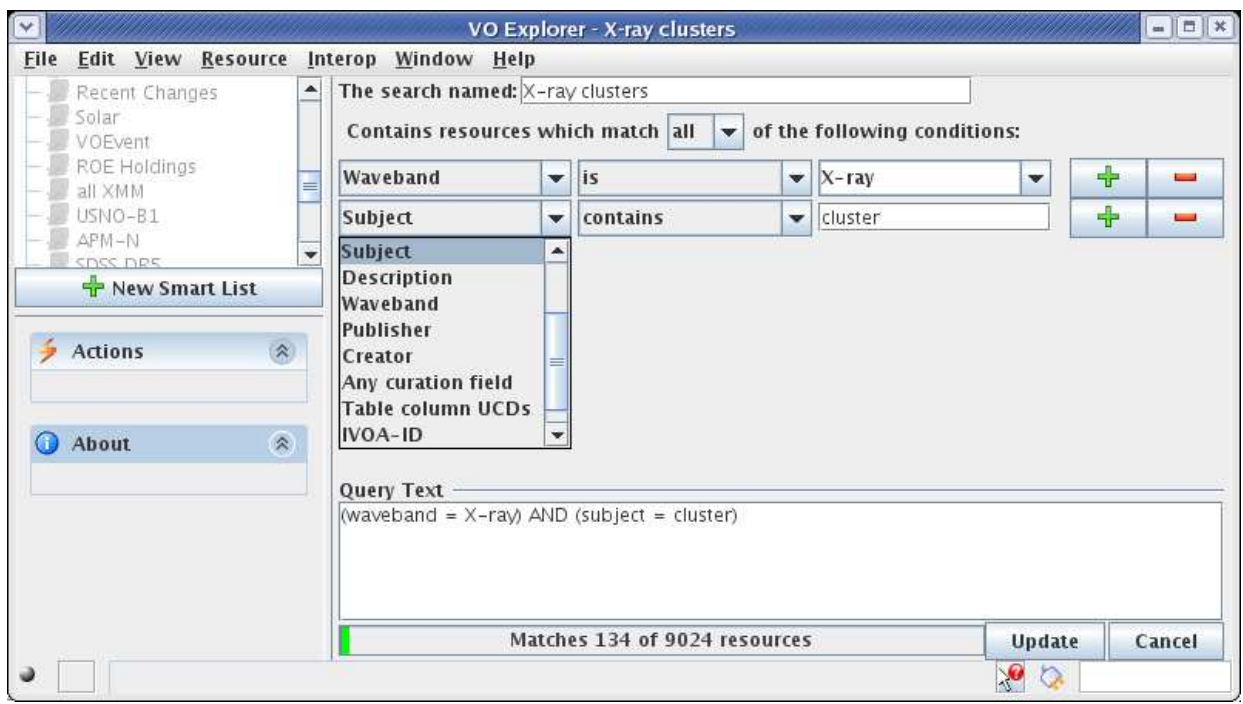}
\caption{The VOExplorer resource search tool (part of the AstroGrid VODesktop software suite): build a list of VO registered resources matching all or any user defined conditions. In this example the user selects ``all'' resources where the waveband is ``X-ray'' and the subject contains the keyword ``cluster'' and 134 matches to these conditions are returned. The user can then filter and analyse the results further via an iTunes type interface as shown in Figure 2.}
\label{fig:1}
\end{figure*}

\begin{figure*}
\centering
\includegraphics[scale=0.7]{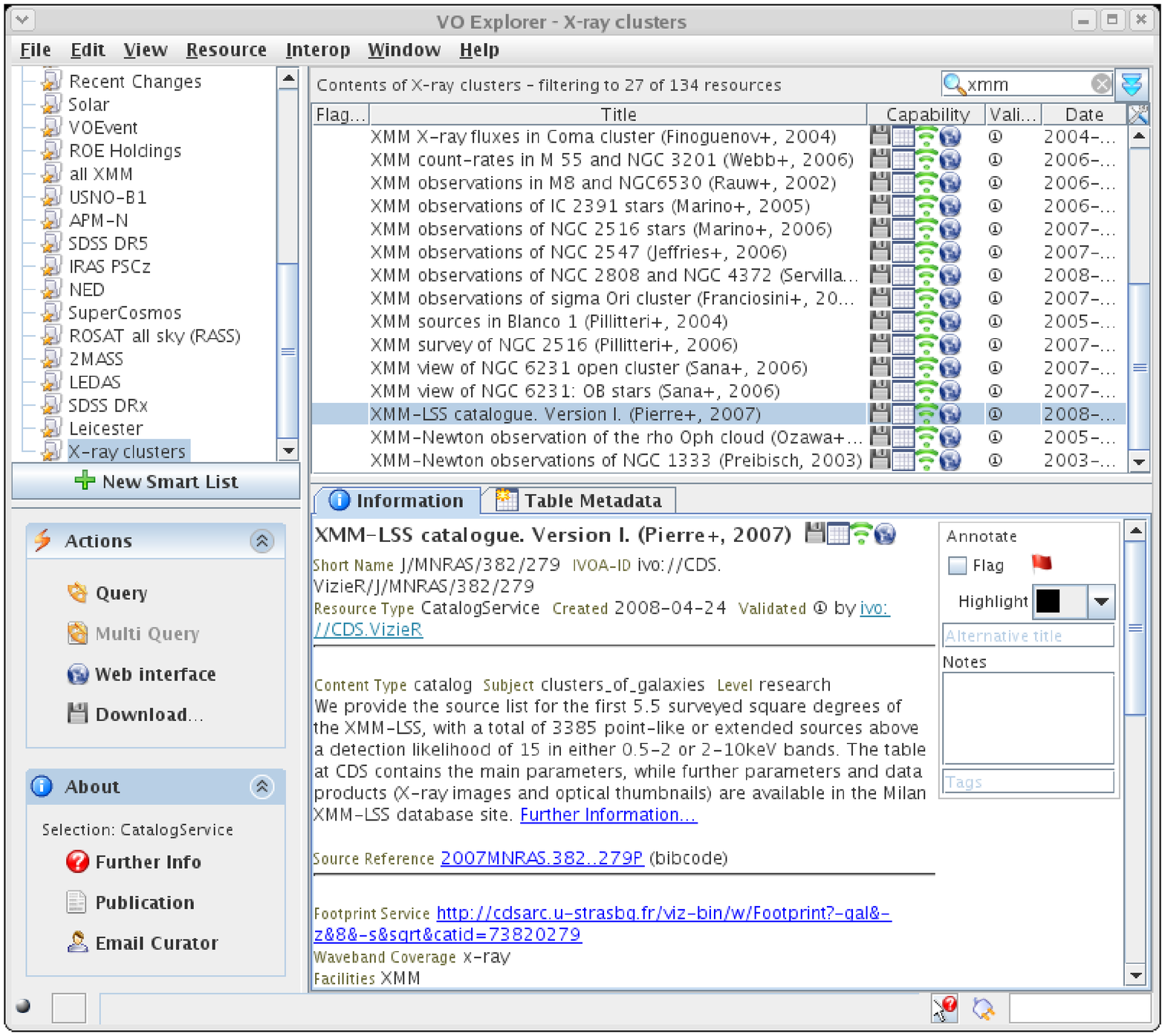}
\caption{Filter VO resources: on listing the resources likely to contain X-ray clusters (Figure 1), the user may filter the metadata by content, coverage or resource type to further refine their search. Here the user has identified ``XMM'' as a search term using the text entry box (top right) leaving 27 resources associated with XMM-Newton satellite data. On selecting one of these resources corresponding to the XMM-LSS catalogue of X-ray clusters of galaxies (Pierre \it{et al.} 2007), the relevant information describing the resource is displayed in the ``Information'' tab and capabilities to query the catalogue are listed under the ``Actions'' pane (left, middle). Positional searches against the catalogue are possible by clicking the ``Query'' or ``Multi Query'' actions for an individual or list of positions respectively as shown in Figure 3. }
\label{fig:2}
\end{figure*}

\section{Science drivers and the advantages of VO access}

For many years one of the primary aims of the Virtual Observatory movement has been to better enable multi wavelength astronomy. Example science drivers include finding all the information on a given set of objects which may be defined by positions, colours or morphology. It is common to study not just the resultant correlations but to identify the so-called ``outlier'' objects, which have unusual properties and which may offer new insight to the astrophysical processes driving the observed emission. Another requirement may be to build a spectral energy distribution from multi archive data while accounting for instrumental, sensitivity and aperture effects. A longer term goal is to compare such correlations on-the-fly with numerical simulations as the latter become available in a more standardised way.

Common aspects to the above use cases include the requirement to browse, search and manage large amounts of distributed heterogeneous data. The astronomer then wants to be able to combine multi wavelength data taking into account differences in units and photometric systems; spatial, wavelength and time coverage; resolution, point spread functions and observing techniques. We set ourselves a challenging target!

The advantages of VO data access in addressing these challenges is to make accessing multi wavelength data easier in that you do not need to access multiple interfaces but can access data from one single entry point. Furthermore the user can then build workflows, i.e. pieces of code which run on the server and are reusable. 

\begin{figure*}
\centering
\includegraphics[scale=0.70]{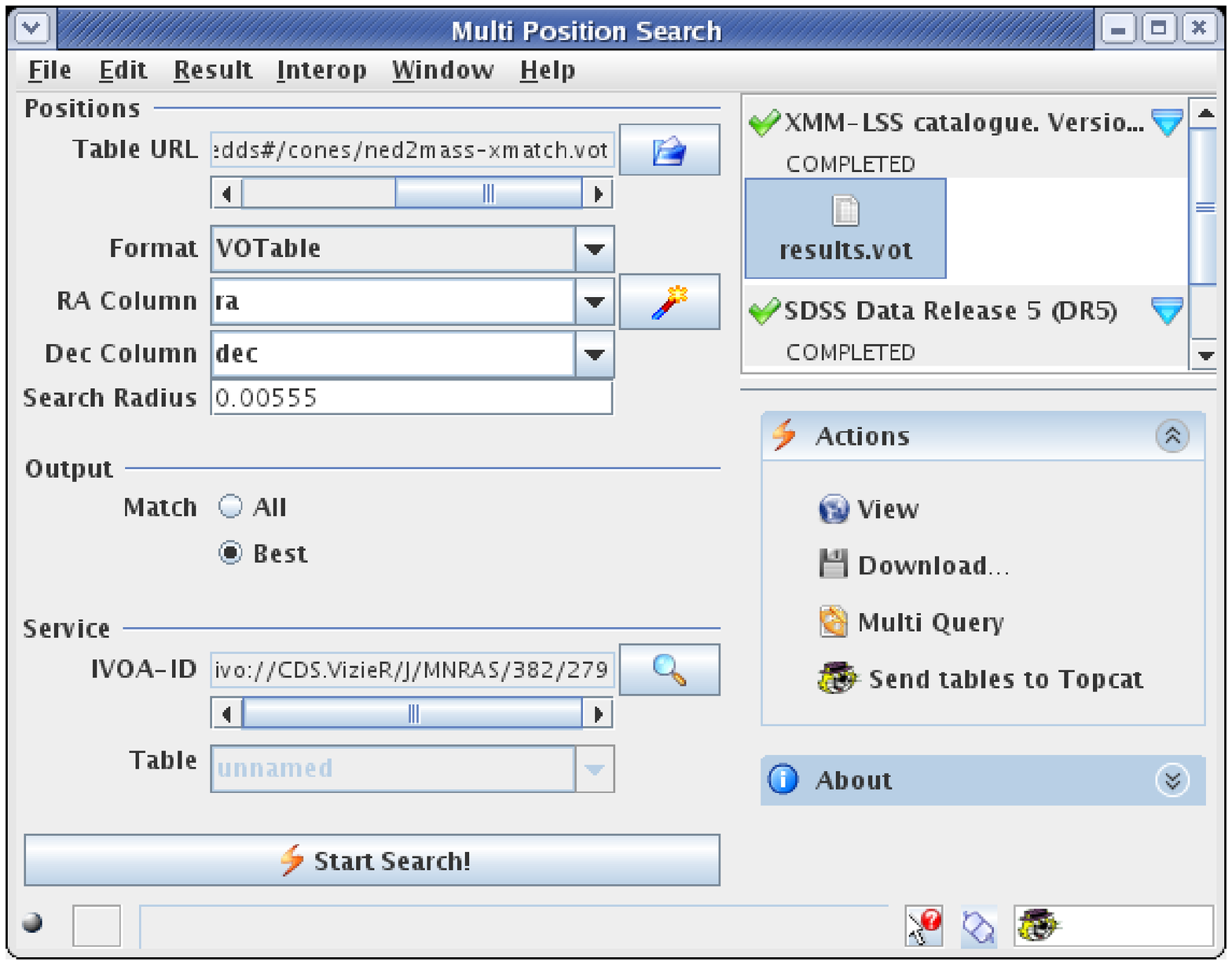}
\caption{Multi position query of science catalogue: on selecting any cone search interface to a catalogue the user may upload their own list of object positions and search against the resource catalogue within a given radius of each object. In this case the user has opted to search the XMM-LSS catalogue of clusters of galaxies (via the Vizier VO interface service selected in Figure 2) and then supplied a list of object positions ``ned2mass-xmatch.vot'' in VOTable format resulting from a cross match of the 2MASS and NED catalogues. In the positions dialogue the format, RA and DEC columns of the list are identified and a search radius. The user has opted to find the nearest (``best'') match in each case in XMM-LSS and on completion of the search a ``results.vot'' table is generated along with actions then possible on the output. If TopCat or other VO enabled tools are running the results can be passed straight to them for further analysis.}
\label{fig:3}
\end{figure*}

\begin{figure*}
\centering
\includegraphics[scale=0.65]{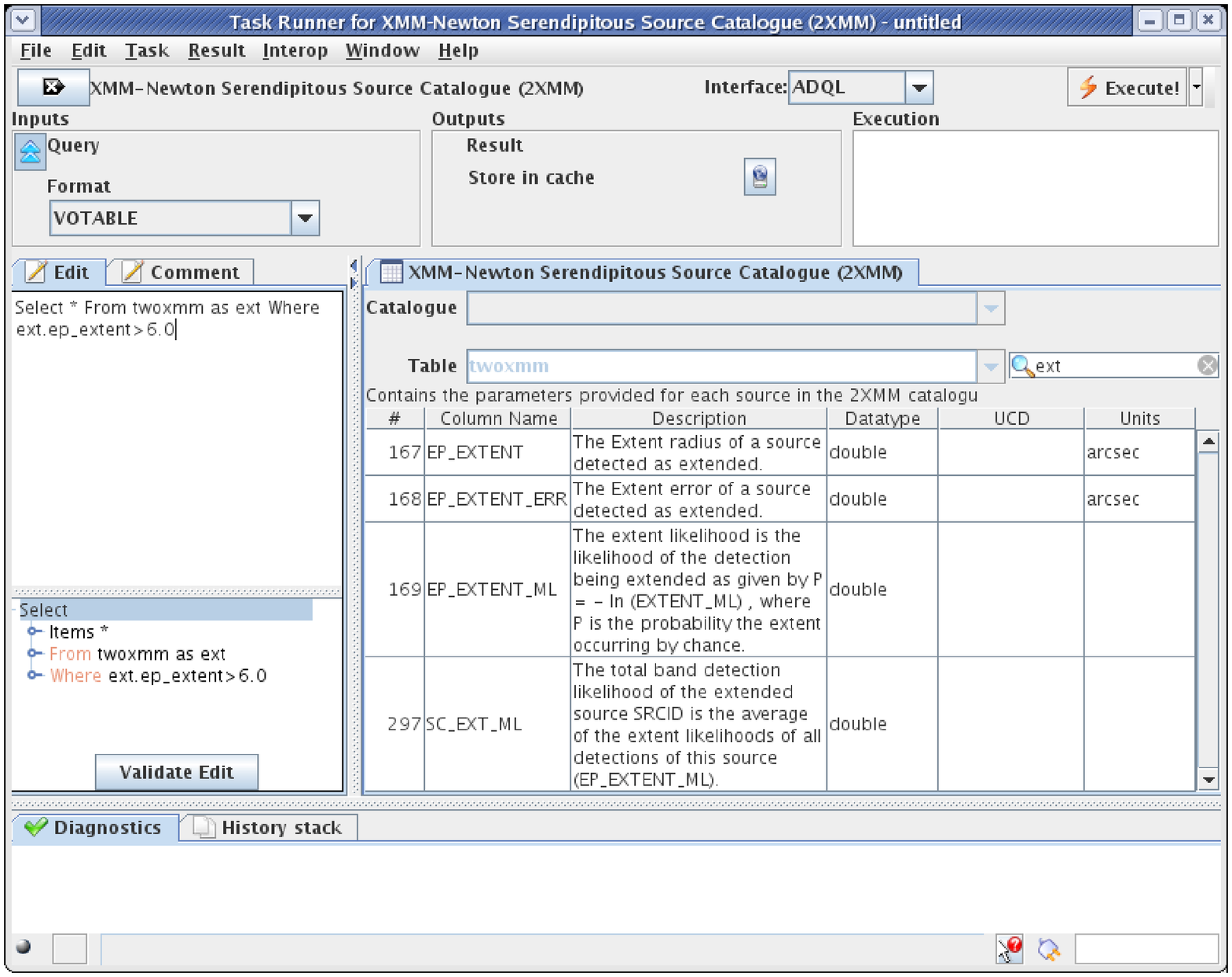}
\caption{Advanced querying including pre-filtering on catalogue parameters: select an ADQL query interface to a 2XMMi X-ray catalogue resource using VOExplorer, then identify columns in the catalogue corresponding to ``extent''. The user may then construct a simple but powerful query against these parameters in the ``Edit'' pane using a standard ``ADQL'' query language and here extract all extended sources (which the user has defined as having extent radius greater than 6.0 arcseconds) to create a new science sample in VOTable format for further visualisation and analysis.}
\label{fig:4}
\end{figure*}

\begin{figure*}
\centering
\includegraphics[scale=1.20]{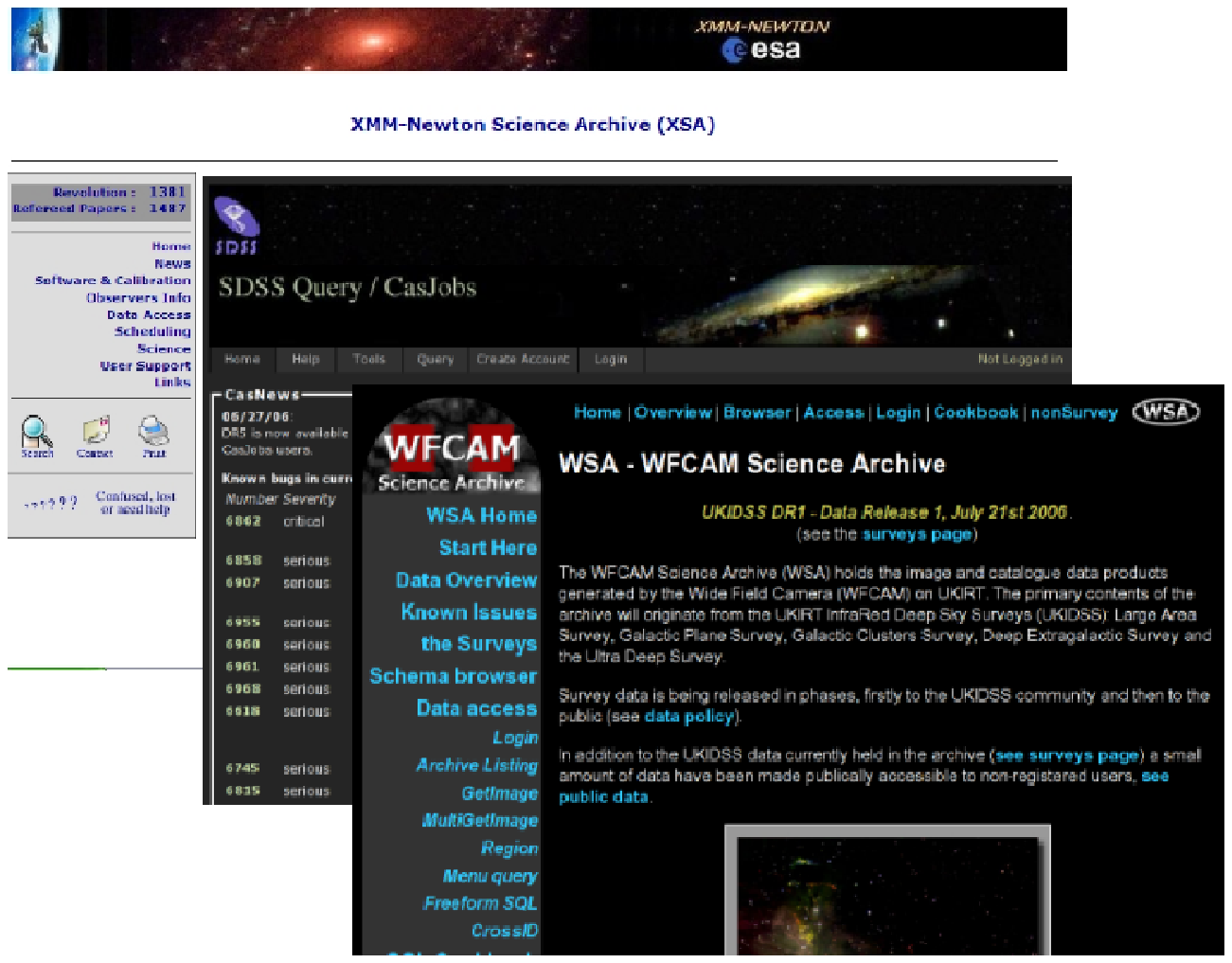}
\caption{Multi mission science archive interfaces: modern astronomers are currently faced with multiple search interfaces for each different mission they wish to search via project webpages. While the expert information provided in each is often essential, each different interface is likely to have it's own terminology and a different flavour of SQL querying. Hence the user ends up searching each one separately and stitching the results back together later. But using the VO to perform standardised ADQL (Figure 4) or more advanced python scripting queries (Figure 6) offers a more efficient solution.}
\label{fig:5}
\end{figure*}

\begin{figure*}
\centering
\includegraphics[scale=0.6]{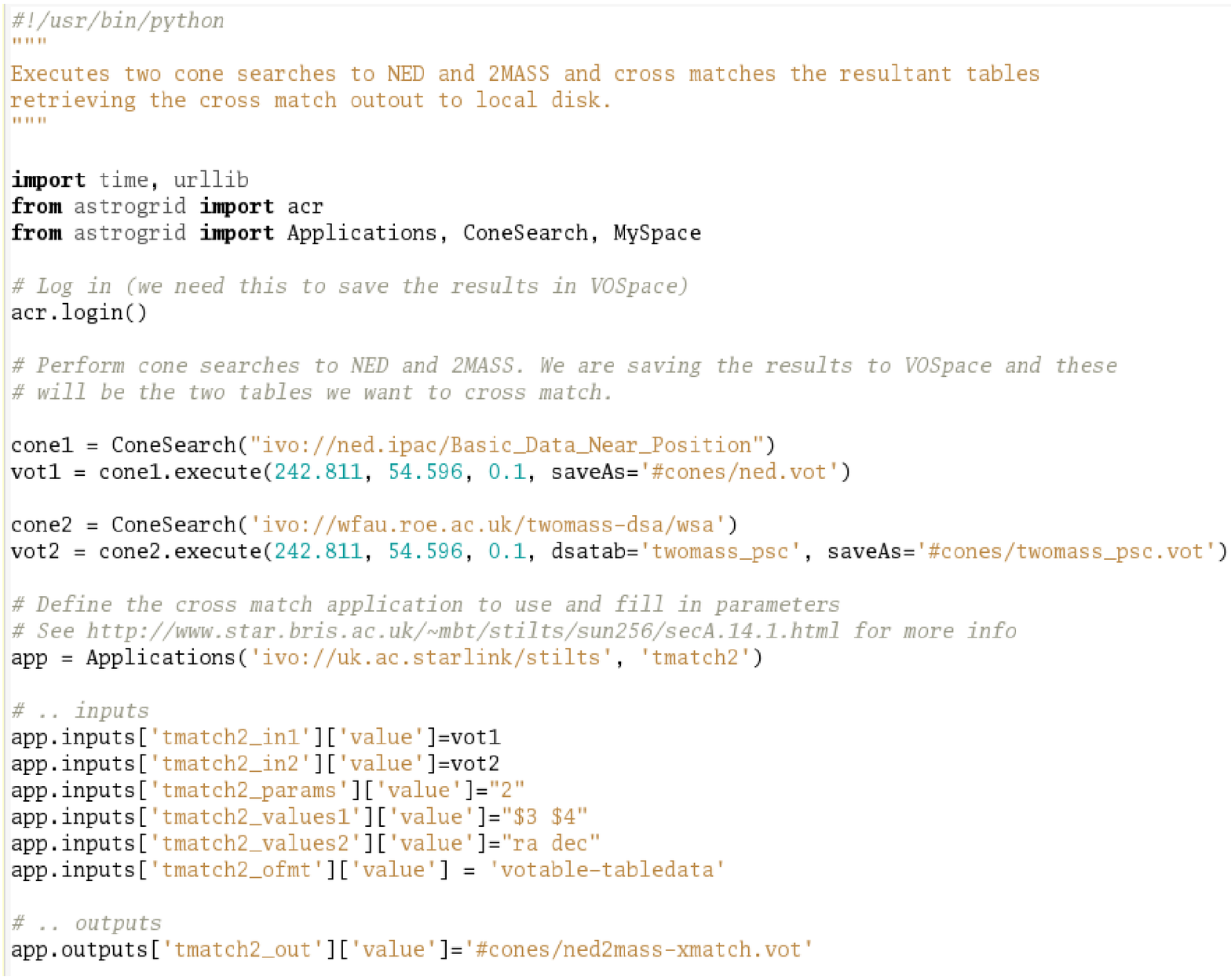}
\caption{VO scripting using python: with the AstroGrid VODesktop running, the user may build advanced scripts in python in order to perform more complex multi mission, multi parameter searches including cross correlation and advanced data mining techniques. In this example the user executes cone searches to NED and 2MASS around a particular object position and then cross matches the resultant outputs before saving to local disk.}
\label{fig:6}
\end{figure*}

\begin{figure*}
\centering
\includegraphics[scale=0.70]{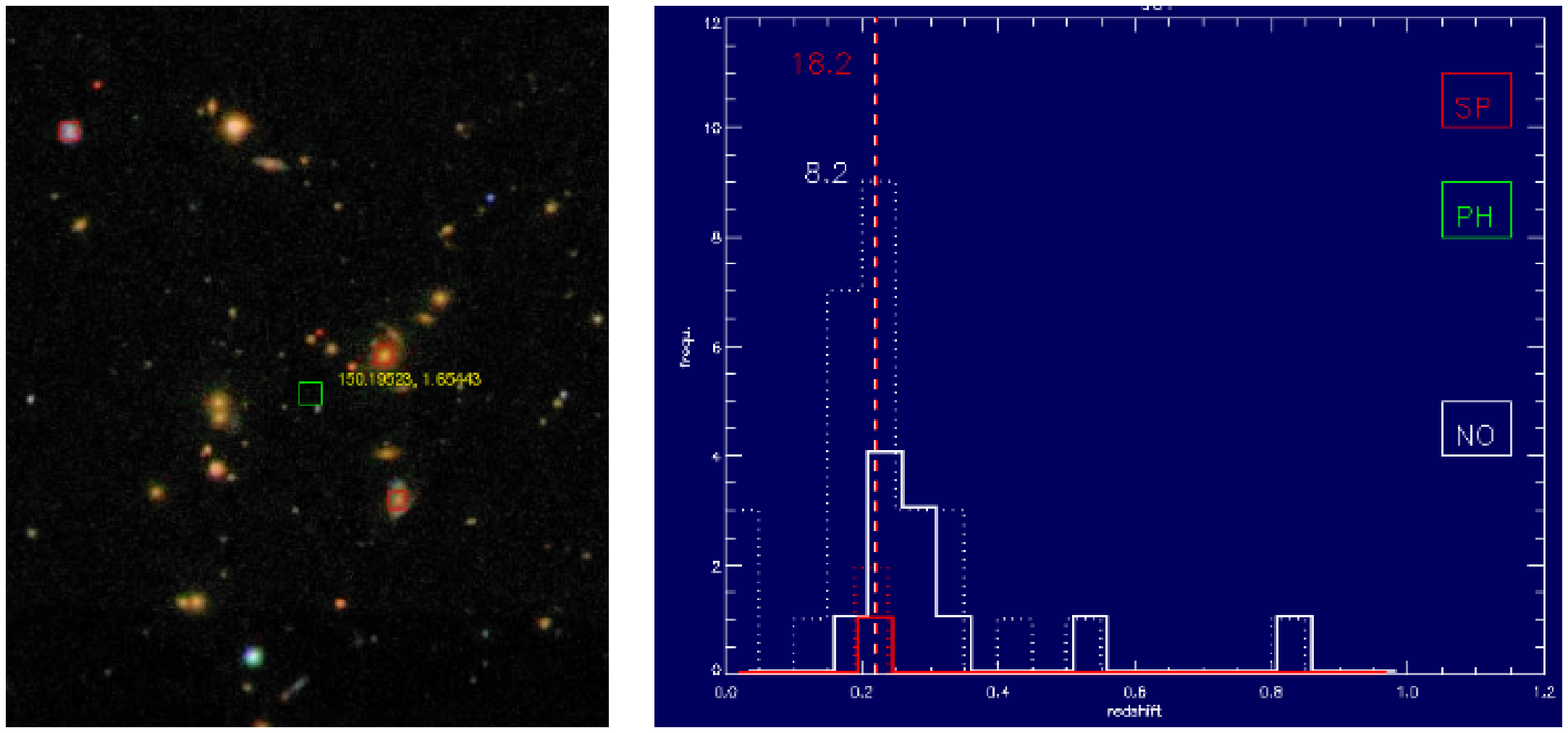}
\caption{An X-ray selected cluster of galaxies candidate from XMM-Newton serendipitous sky survey (Watson et~al 2009), selected as shown in Figure 4, then correlated with the Sloan Digital Sky Survey DR5 optical photometric catalogue (Adelman-McCarthy et~al 2008), as imaged in left panel, to extract an averaged redshift (right panel) of z=0.223 (Lamer et~al, in prep.).}
\label{fig:7}
\end{figure*}

The range and complexity of datasets and resources published to the VO is rapidly increasing. These resources are heterogeneous -
and are published through various standard interfaces allowing access
to images, catalogues, spectra, transient event data, tool interfaces and so on. Providing access to these resources is a success of the VO
movement, where effective use of newly emerging publishing standards
as provided by the International Virtual Observatory Alliance
(IVOA, http://www.ivoa.net) has been made by the astronomy community. Information about each
resource published to the VO is entered into a top level,
continually updating ``registry'', this providing in effect a record of where and
what the resource is.

With the advent of these many resources available through the VO,
an emerging challenge is how to offer the astronomer a reliable and
usable means to search, retrieve and visualise the relevant data and
resources to meet the needs of their particular science problem.

\section{Introducing the AstroGrid VO Desktop}

In April 2008, the UK AstroGrid VO project made the first public release of the VO Desktop suite of applications which consists of several interlinked tools including the VOExplorer resource browser, File Explorer, Task Runner and Query Builder. You can search for resources and data in the VO; bookmark your favourites; fetch images, spectra, light curves and catalogues; run queries on databases; save and share files in VOSpace; and invoke remote applications. VO Desktop also runs background software called ``Astro Runtime'' which handles all interactions with the remote services. 

VOExplorer offers a powerful data-centric visualisation for browsing and filtering the
entire VO registry using a familiar ``iTunes'' type interface. The Registry stores information about Resources all round the world. A ``Resource'' could mean a data collection like UKIDSS, or an application like Sextractor that you can invoke as a remote service, or just information about an organisation. Data collection resources will usually have one or more ``capabilities'', i.e. ways of accessing the data, like an image cut out service, or a catalogue conesearch, or a full query-language (ADQL) interface. Each registry entry contains metadata describing the resource - basically a set of standard attribute=value pairs. This information will tell you whether the resource is a catalogue or an image atlas, whether it has infrared or X-ray data, as well as who curates the data. To locate particular object(s) of interest, say, you need to send a query to the resource itself. Note that the Registry is maintained by AstroGrid, but the information in it is provided by the resource owners, not by the AstroGrid project. This allows you to bookmark
your own personalised lists of resources for repeat use, filter using any available metadata and then to run tasks on the selected resources as desired.

We now illustrate the use of the AstroGrid software to tackle a typical multi wavelength science case involving X-ray selected clusters of galaxies at different levels of complexity and show how data or search and correlation results may be saved to remote or local storage as well as passed to any suitably enabled VO tool for further visualisation and analysis.

\section{VO Desktop in action: X-ray clusters of galaxies}

In Fig 1 we show a screenshot of the VOExplorer resource search interface which allows a user to build a set of AND, OR conditions. This example is a simple search for resources containing both ``X-ray'' in the waveband AND requiring the subject contains the word ``cluster''. A list of resources is then returned and the search may now be refined further by means of metadata filter wheels acting on content, coverage or other resource type. In Fig 2 the resultant resources that satisfy this condition are displayed following a further filtering to any resource including the keyword ``XMM'' denoting the XMM-Newton satellite. On selecting the XMM-LSS catalogue of X-ray selected clusters of galaxies (Pierre et~al 2007), relevant ``Actions'' available for the selected resource are listed including a positional search facility ``Query'' which launches the AstroScope search tool and also a ``Multi Query'' may be launched for a list of positions to be supplied by the user (Figure 3). The output can then be passed directly into any VO tool also running such as TopCat or saved to disk. Similarly for solar datasets or transient event data one can launch the HelioScope or VOScope tool on selected datasets to search by time interval.

As an example of a more advanced task, Figure 4 illustrates how VOExplorer can be used to identify an ADQL (a standard Astronomical Data Query Language,  based on a simple subset of SQL) query interface to the recent XMM-Newton X-ray serendipitous survey catalogue, 2XMM (http://xmmssc-www.star.le.ac.uk/Catalogue/2XMM/, Watson et~al 2009), available in the VO registry alongside thumbnail image and other related products via LEDAS (LEicester Database and Archive Service, http://ledas-www.star.le.ac.uk) or in more basic form via ESA and CDS archives. The user can then filter the available 2XMM columns and set conditions on suitable extent parameters and then perform a simple 3-line ADQL query to select and return all extended X-ray sources (column ep\_extent $\geq$6 arcsecs) in the catalogue to standard VOTable or other format. 

\section{More advanced querying and data mining: cross matching 2XMM and SDSS}

Typically, the modern multi wavelength astronomer has to access many different mission archives in order to select samples from each using a number of individual project webpages (Figure 5). Of course the expert information provided for each catalogue or data product for each differnet mission under study is usually essential in order to fully understand the criteria before a full scientific analysis. But each different interface is also likely to have it's own terminology and a particular flavour of SQL querying also. Hence even to perform simple cross matching involves searching each one separately and then stitching back together each of the  outputs afterwards. This can be a non-trivial task.

However, ADQL searches and in particular more advanced cross matching to other archives as well as other complex tasks may now be performed using a common schema in python with AstroGrid. The user may run commands directly from the command line as long as the AstroGrid Runtime capability is running in the background. Figure 6 shows a reasonably simple example chosen from a number of template python scripts for the VO, as provided by AstroGrid. In it, the user executes cone searches to NED and 2MASS around a particular individual object position and then cross matches the resultant outputs before saving to local disk. Now, by selecting an X-ray extended source sample from the 2XMM serendipitous source catalogue, as described previously using ADQL or python scripting, and then correlating the results with the Sloan Digital Sky Survey (SDSS) optical catalogues, e.g. Adelman-McCarthy et~al (2008), one could derive an averaged photometric redshift for each X-ray selected galaxy cluster candidate and then repeat for the entire sample and compare to any catalogued spectroscopic redshifts. Figure 7 shows an example of the fruits of this kind of work for one such XMM cluster candidate by Lamer et~al (in prep) and now made possible for large samples using the VO in a fraction of the time. 

Finally, once a science sample is made the user is offered the opportunity to process and act
on those data sets. For this purpose a range of data visualisation and
analytical tools, often building on previously existing tools developed over many years, are available via the PLASTIC/SAMP messaging protocol. Popular tools include TopCat for catalogues, Aladin and Gaia for images, SPLAT and VOSpec for spectra. These are described in more detail elsewhere in these proceedings.

\section*{Acknowledgements}

The author wishes to acknowledge the superb team effort by everyone involved in the UK AstroGrid project, working in partnership with our European-VO and IVOA colleagues. This research has made use of data obtained and software provided by AstroGrid, which is funded by the UK Science and Technology Facilities Council and through the EU's Framework 6 and 7 programmes. We also acknowledge the contributions to this work by the XMM-Newton Survey Science Centre funded by STFC and European Space Agency partners.



\begin{thebibliography}{}

\bibitem{}
  Pierre M., et~al. 2007, MNRAS 382, 279

\bibitem{}
  Watson M.G., et~al. 2009, A\&A 493, 339
  
\bibitem{}
  Adelman-McCarthy J.K., et~al. 2008, ApJS 175, 297  

\end{thebibliography}
\end{document}